\documentclass[jcs]{iosart2x}
\usepackage{amsmath}
\usepackage{amssymb}
\usepackage{multirow}
\usepackage{mathtools}
\usepackage{graphicx}
\usepackage{xcolor}
\usepackage{oubraces}
\usepackage{bm}
\usepackage{algorithmic}
\usepackage{subfig}

\DeclarePairedDelimiter\floor{\lfloor}{\rfloor}
\definecolor{mygray}{gray}{0.8}
\newcommand{\highlight}[1]{\colorbox{mygray}{$\displaystyle #1$}}
\begin{document}
\makeatletter
\let\put@numberlines@box\relax
\makeatother
\begin{frontmatter}


%
%

\title{Securing LSB embedding against structural steganalysis}
\runtitle{Securing LSB embedding against structural steganalysis}
\author{Brian~A.~Powell}
\address{The Johns Hopkins University Applied Physics Laboratory, Laurel, MD 20723}

\maketitle

\begin{abstract}
This work explores the extent to which LSB embedding can be made secure against structural steganalysis through a modification of cover image statistics prior to message embedding.  LSB embedding disturbs the statistics of consecutive k-tuples of pixels, and a $k^{\rm th}$-order structural attack detects hidden messages with lengths in proportion to the size of the imbalance amongst sets of $k$-tuples. To protect against $k^{\rm th}$-order structural attacks, cover modifications involve the redistribution of $k$-tuples among the different sets so that symmetries of the cover image are broken, then repaired through the act of LSB embedding so that the stego image bears the statistics of the original cover.  We find this is only feasible for securing against up to $3^{\rm rd}$-order attacks since higher-order protections result in virtually zero embedding capacities.  To protect against 3rd-order attacks, we perform a redistribution of triplets that also preserves the statistics of pairs. This is done by embedding into only certain pixels of each sextuplet, constraining the maximum embedding rate to be $\leq 2/3$ bits per channel.  Testing on a variety of image formats, we report best performance for JPEG-compressed images with a mean maximum embedding rate undetectable by $2^{\rm nd}$- and $3^{\rm rd}$-order attacks of 0.21 bpc.
\end{abstract}

\end{frontmatter}


\section{Introduction}
\begin{figure}
\centering
\includegraphics[width=4in]{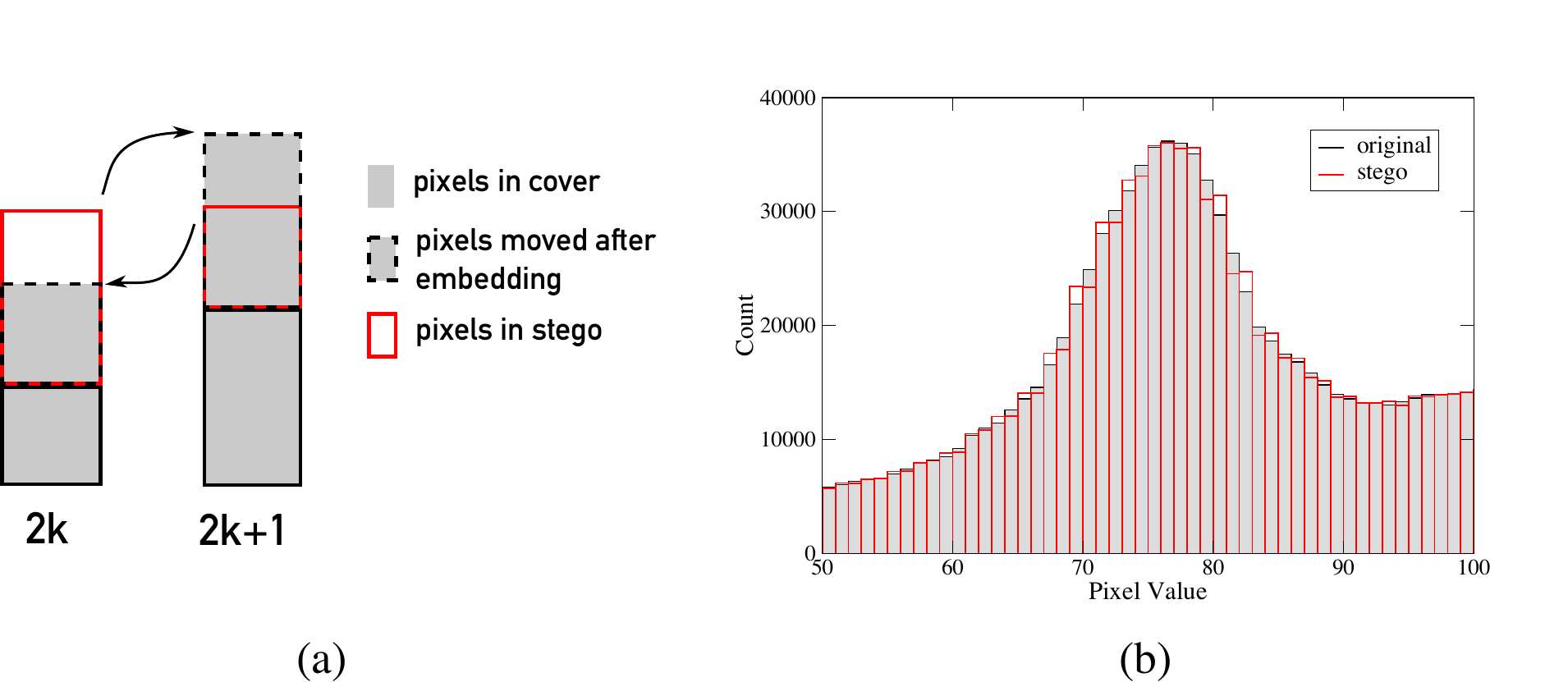}
\caption{\footnotesize{(a) Illustration of how pixel value pairs $(2k,2k+1)$ even-out under LSB embedding. (b) Histogram of pixel values comparing original image (gray) with stego image (red) after complete LSB embedding. Notice how neighboring pixel values tend to equality.}}
\label{hist_attack}
\end{figure}

Hiding secret messages in the least significant bits of pixels in digital images is the oldest steganographic technique.  It follows a simple rule: to embed a message bit into a pixel of value $x$, flip the pixel's least significant bit (LSB) to match the message bit,  
\begin{equation}
{\rm flip}(x) = \left\{\begin{array}{ll}
x+1\,\,{\rm for }\,\, x\,\, {\rm even} \\
x-1\,\,{\rm for }\,\, x\,\, {\rm odd}. \end{array}\right.
\label{LSBflip}
\end{equation}
The beauty of this technique is the simplicity of message retrieval: one needs merely to read off the LSBs of the pixels (perhaps scrambled in some way) to obtain the hidden message.  No special software or complex operations are needed. Since only the LSBs of only some pixels are modified, LSB embedding is also virtually impossible to detect visually.  Alas, it does tend to affect pixel value statistics in an idiosyncratic way. Consider a pair of consecutive pixel values, $2k$ and $2k+1$.  If we embed a message bit of \texttt{1} into the pixel of value $2k$, its value becomes $2k+1$.  Conversely, if we embed a message bit of \texttt{0} into the pixel of value $2k+1$, its value becomes $2k$.  Meanwhile, embedding a \texttt{0} into the $2k$ pixel or a \texttt{1} into the $2k+1$ pixel results in no change.  If we have an equal chance of embedding into a $2k$- or $2k+1$-valued pixel ({\it i.e.} if the cover image has nearly an equal number of even- and odd-valued pixels), and if the message bit has a 50\% chance of being either a \texttt{1} or \texttt{0} ({\it e.g.} if the message is encrypted and so pseudo-random), then LSB embedding tends to ``even out'' consecutive pairs of pixel values $(2k,2k+1)$, converting even-valued pixels to odd and odd-valued pixels to even in approximately equal numbers. The extent to which these even out depends on the stego load: if we fully embed the cover the image, then we expect the number of pixels of value $2k$ and $2k+1$ to become nearly equal, $n_{2k} \approx n_{2k+1}$.  Figure \ref{hist_attack} shows the effect on pixel values of a fully embedded image.  

This asymmetry has given rise to a barrage of steganalytic attacks over the last two decades, starting in 2000 with the histogram attack \cite{Westfeld1} which seeks to detect LSB embedding by examining the extent to which neighboring bins in the image's pixel value histogram tend to equality as a result of this embedding asymmetry. This attack only works well for high embedding rates, and so new techniques looking for changes in higher-order statistics, like correlations among neighboring pixel values, were developed \cite{Fridrich1,Dumitrescu1,Dumitrescu2,Dumitrescu3,Ker1,Ker2,Ker4}.  These powerful methods are generally referred to as {\it structural steganalysis}, because they target the statistical properties of image structures (like pixel pairs, triplets, and so on).  Structural steganalysis is based on the idea that the cardinalities of certain sets of consecutive pixel groups should be approximately equal for natural images, but diverge in an idiosyncratic way under LSB embedding.  Structural attacks analyze the count statistics of these pixel groups and render an estimate of the hidden message length in proportion to this divergence. The most sensitive attacks in this family are able to detect embedding rates as low as 3\% \cite{Ker1}, and are more accurate than other prominent attacks against LSB embedding \cite{Dabeer,Fillatre,Zhang,Fridrich2,Ker5} for certain image formats, like JPEG-compressed images. We discuss these methods at length in Section 4.

In this work, we seek a means of securing LSB embedding against higher-order structural steganalysis, including both RS/Sample Pairs Analysis (SPA) \cite{Fridrich1,Dumitrescu1,Dumitrescu2} and Triples analysis \cite{Ker2}, that 1) does not produce obvious statistical artifacts, 2) does not require significant additional secret data for message recovery, and 3) preserves the computational and algorithmic simplicity of LSB embedding.  We develop a method of {\it cover modification},  in which pixel values of the cover are modified, disturbing the natural count statistics of various pixel groups prior to message embedding, such that the act of embedding the message restores the count statistics to their natural values.   With apparently normal set cardinalities, structural analyzers will then be fooled into concluding that no hidden message is present.  This kind of cover modification was presented as a defense against SPA in \cite{Shreelekshmi}, and we extend it here to protect against up to third-order (Triples) attacks.  This extension is not straight-forward: to protect against all orders up to some order $k$, the statistics of $n$-tuples must be preserved, where $n$ is the least common multiple of all orders up to and including  $k$.  We find that cover modifications in terms of sextuplets, required to preserve the statistics of both pairs and triplets, are highly constrained and result in cover images with virtually zero embedding capacity.  We demonstrate how to instead perform cover modifications at third-order, which is much less constrained, by redistributing triplets in such a way that also preserves the statistics of pairs.  The trade-off is that only certain pixels in the image can be embedded, and so steganographic capacities are reduced.

We test this method against a range of different image types and find that we can achieve maximum undetectable embedding rates of 0.12, 0.17, and 0.21 bits per channel for uncompressed grayscale, uncompressed color, and JPEG-compressed color raster images.  We also argue that extending protections to higher-order (quadruples analysis \cite{Ker3}) is only possible at the cost of virtually zero embedding capacity; however, such detectors are difficult to implement in practice.

This paper is organized as follows: Section II explores related prior art on the subject of improving the security of LSB embedding against structural steganalysis, and Section III provides some rationale for still considering LSB embedding in today's steganographic landscape. Section IV reviews the family of structural steganalysis techniques and in Section V we introduce the procedure of cover modification as it is used to secure LSB embedding against SPA in \cite{Shreelekshmi}, with some new elements necessary for extending it to higher-order.  Section VI develops the new cover modification procedure to protect against both second- and third-order structural attacks, and presents results of testing on a data set of grayscale and color images.  In Section VII we discuss the possibility of extending this methodology to higher-order, and in Section VIII we conclude.
\section{Related Work}
Due its simplicity, there has been much work on improving the security of LSB embedding against the ever-escalating wave of steganalytic attacks. These approaches evade statistical attacks focused on image structures by preserving these statistics during the embedding process.  There are three broad approaches to this problem: {\it embedding strategies}, in which LSB embedding is only performed on subgroups of pixels that preserve certain statistics;  {\it statistical restoration}, in which portions of the cover image are altered after LSB embedding to recover certain statistics of the cover image; and {\it cover modification}, in which the cover image is altered prior to embedding so that the stego image retains certain statistics of the cover image.  Some methods incorporate more than one of these aspects.  We review several relevant works in this section. 
 
The earliest approaches attempted to circumvent histogram-based attacks.  In the work of \cite{Franz}, the histogram is preserved by encoding the message such that the probabilities of 1's and 0's in the message are precisely those required to keep the frequencies of adjacent bins unchanged.  Some protection against second-order statistics is conferred if the pixel pairs are chosen such that their frequencies in the image co-occurrence matrix are unchanged after the embedding; however, this method is not protective against second-order attacks like RS analysis \cite{Bohme}.  This procedure is only applicable for high embedding rates, since otherwise the histogram attack is not particularly effective.  In \cite{Eggers}, the {\it histogram-preserving data mapping} introduced under the assumption that pixels are i.i.d., embeds data with the same distribution as the cover image histogram so as to minimize their relative entropy.  The i.i.d. assumption is in general not true for natural images, however, and so this method is susceptible to higher-order structural attacks, {\it e.g.} as shown in \cite{Tzschoppe}.

The LSB+ method of Wu {\it et al.} \cite{Wu} also seeks to preserve the pixel value histogram by compensating for bits embedded into a given pair of neighboring bins by appropriately changing the values of other pixels from these bins reserved for this purpose.  To protect against second-order attacks, like SPA, only restricted groupings of pixels are embedded so that these statistics too can be preserved; the result is that test images had a very low average embedding capacity of around 2.5\%.  Protection against higher-order statistics would result in even lower embedding capacities.  An improvement in capacity is offered by \cite{Ghazanfari} but this method is equally susceptible to higher-order structural steganalysis.   
 
With the increasing use of the powerful SPA technique, steganographers recognized the need to go beyond the preservation of first-order image statistics.  In \cite{Marcal}, an inverse histogram transformation is applied to the cover image prior to embedding.  This operation compresses the range of pixel values in the cover image, essentially coarse-graining the image prior to embedding. Since structural attacks like RS and SPA rely primarily on trace sets with small differences among neighboring pixels, as these are the most common pairs, the method of \cite{Marcal} is able to defeat these attacks for a wide range of embedding rates.  To recover the hidden message, the recipient must reverse the compression.  The trouble with this method is that the compression transformation eliminates entire pixel values from the stego image, which appear as empty bins in the image histogram.  Analysis of the histogram enables one to reverse the transformation and then directly analyze the LSB-embedded image. Lou and Hu \cite{Lou} correct this problem by performing multiple transformations with different parameters on different pixel groups of the cover, with the result that the combined histogram has no missing levels.  Depending on the pixel grouping strategy, the authors of \cite{Lou} acknowledge that this approach could be susceptible to a brute force attack wherein the steganalyst examines the histograms of many different pixel groups looking for evidence of missing levels.  As the number of pixel groups grows, such that the histograms contain ever fewer pixels, missing levels could occur naturally and the authors argue that in this case there are insufficient histogram statistics to support steganalysis.  This claim, however, remains to be validated in general. 

An interesting example of statistical restoration is provided by the method of {\it dynamic compensation} \cite{Luo}.  Here the message is embedded and the values of half of the image pixels are increased by one.  This has the effect of essentially ``resetting'' the statistics of the stego image, and structural steganalysis is unable to detect any hidden messages.  The pixels used for this compensation must generally be chosen dynamically such that detection by common structural attacks is minimized. The main drawback of this method is that message retrieval requires a reversal of this compensation procedure, and so the locations of all modified pixels must be communicated to the recipient.  This is a sizable amount of data: for a 512x512 image, this amounts to a 300kB secret key that must be securely exchanged along with the image.
A related approach was explored in \cite{Shreelekshmi1} where half of the image LSBs are flipped after embedding so that SPA and RS tests are fooled into concluding that images are maximally embedded regardless of the true embedding rate.  While indeed these tests are in error, it is not clear how this result safeguards the stego image since such a detection would likely arouse suspicion that the image was either fully LSB embedded or had at least been tampered with.  Further, message extraction requires knowledge of which LSBs were flipped so that the this operation can be reversed.  

An approach that combines cover modification with an embedding scheme based on the eight-queens problem is presented in \cite{Bansal}.  Here, each LSB is flipped or not according to whether its pixel, when taken as part of an eight-pixel block, is masked by one of the 92 eight-queens solutions.  A group of pixels is reserved to restore set cardinalities to approximate those of the cover so that SPA is unable to detect the message.  A general upper bound on embedding capacities is not established in \cite{Bansal}, but sample images are tested up to relative payloads of 30\%.  This is lower than the cover modification technique of \cite{Shreelekshmi}, discussed below, and similarly does not protect against higher-order attacks.  It should also be noted that message extraction using eight-queens encoding is greatly more complicated than simple LSB embedding, and does not confer additional security against structural steganalysis.

Most recently, the work of \cite{Shreelekshmi} considers cover modification where the cardinalities of sets analyzed by SPA to detect the presence of LSB steganography are adjusted prior to message embedding such that the relevant second-order statistics are preserved in the process.  This approach successfully protects against SPA at the cost of lower embedding capacities, upwards to around 50\% on average \cite{Shreelekshmi}.  Though second-order statistics are carefully preserved in this method, higher-order statistics can still be targeted by Triples analysis to uncover the hidden message length.  

\section{Why LSB Embedding?}
The technique of substituting, or embedding, message bits into the least significant bits of cover image pixels is perhaps the oldest and arguably the simplest steganographic technique.  Since its inception, LSB embedding has been targeted by a wide range of steganalysis, and is considered today to be effectively broken.  A number of techniques were soon developed that incorporate {\it LSB matching} \cite{Sharp} into more secure frameworks\footnote{In LSB matching pixel values are randomly changed by $\pm 1$ so that their LSB's match message bits; this process does not cause obvious statistical artifacts like LSB embedding.}.  These include encoding schemes \cite{Soukal,Mielikainen,Filler,Filler2} to reduce the number of modified pixels, adaptive embedding strategies \cite{Wu2,Luo2,Pevny,Holub,Holub2,Fridrich3,MiPOD,HILL,LiaoX} which select pixels for modification that minimize some measure of distortion, and methods that employ game-theoretic optimizations \cite{GT1,GT2}.  The adaptive strategy, HILL \cite{HILL}, was one of the most successful algorithms on the BOSS image database \cite{Chaumont} as of 2016.  Since that time, machine-learning-based methods have come to the fore \cite{GSIVAT,SSGAN,ASDL-GAN,DCGAN,HiDDeN,TangW,ADV-EMB,UT-GAN,LiuJ}, some of the most powerful using generative adversarial networks (GANs) to embed images in ways undetectable by prospective deep learning-based steganalyzers.  Given this prodigious improvement in the state-of-the-art, one would expect steganagraphy based on LSB-embedding to be effectively extinct.  And, yet, LSB embedding is still implemented in a large number of open source and commercial data hiding products \cite{Fridrich4}.  Even within academia, there is considerable research interest: at least a dozen papers in 2021 alone, found via a \texttt{scholar.google.com} database search, focused on various applications and security improvements of LSB embedding. 

Possible reasons for the tenacious popularity of LSB embedding include 1) its simplicity (in terms of code, compute, storage, and hardware requirements), 2) its availability, 3) its ease of message extraction (in terms of additional data required, beyond perhaps a once-pre-shared key), and 4) that the expected threat of sophisticated steganalysis is not sufficiently high to warrant more advanced approaches.  While it is difficult to know to what extent the use of steganography over lower-risk channels influences the popularity of LSB embedding, pragmatism argues that one ``do only what is necessary, and no more.''  This maxim shapes standard tradecraft in fields like penetration testing, in which simple, unsophisticated attacks are used whenever and wherever possible.  We argue below that state-of-the-art methods suffer from a number of these kinds of impracticality, favoring the use of simpler stegangraphy, like LSB embedding, particularly over lower-risk channels.  In this paper, we therefore seek to improve its security against those steganalytic attacks designed to target it in practice.  

\subsection{Simplicity and Availability}
To exemplify the simplicity of LSB embedding, it can be implemented with an 80-character Perl code at the Linux command-line \cite{Ker1}.  Software for both message embedding and extraction are widely available for free on the Internet (see \cite{ListWiki} for a list of steganography programs, many of which are free and include LSB embedding). The code used to perform the cover modifications and embedding described in this paper is publicly available\footnote{\texttt{https://github.com/bapowellphys/LSB\_cover\_mods}}: it is written in Python and makes use of standard libraries.  This code takes a few seconds to create a 512x512 stego image on a MacBook Pro with 2.3 GHz processor and 16 GB RAM.  While the algorithmic complexities of adaptive methods like HUGO \cite{Pevny}, UNIWARD \cite{Holub2}, and HILL \cite{HILL} are considerably greater than LSB embedding, these techniques are not particularly resource intensive and can be run on standard hardware.  Much of this software is also available in C/C++ and Matlab \cite{FridrichCode}.  

In contrast, methods employing machine learning are significantly more complex and pre-trained networks are generally unavailable; for example, several popular implementations \cite{SSGAN,ASDL-GAN,TangW,UT-GAN} have no reported open source code.  These techniques require considerable expertise to develop from scratch: as examples typical of this class of methods, the works \cite{ASDL-GAN,HiDDeN,TangW} make use of sets of deep convolutional neural networks that must be trained via adversarial learning.  Generally tens or hundreds of thousands of images \cite{GSIVAT,DCGAN,ASDL-GAN,HiDDeN,ADV-EMB} are required for training, ideally on high-performance hardware like GPUs to speed-up training and hyperparameter optimization.  Training times vary, ranging from upwards of 78 hours for ASDL-GAN \cite{ASDL-GAN} to 9 hours for UT-GAN \cite{UT-GAN} on a single GPU.  These methods, while state-of-the-art, are very much research-oriented and not suited for wide deployment outside of academia. In short, actors employing steganography to send secret messages are generally not artificial intelligence engineers capable of training deep neural networks.  

A further observation is that, while some GAN-based methods have embedding capacities competitive with state-of-the-art adaptive techniques \cite{steganogan}, most can accomplish at most 0.4 bits per pixel with detection error rates in the 20\%-30\% range \cite{GSIVAT,ASDL-GAN,HiDDeN,ADV-EMB,UT-GAN}, and are limited to working with smaller images (32x32 in the case of \cite{GSIVAT}) or image patches (16x16 for \cite{HiDDeN}).   
\subsection{Ease of Message Extraction}
Traditional steganography embeds encrypted messages in the cover image; this is done both for security and because the resulting pseudo-random bit stream nicely randomizes pixel modifications. Messages are also typically embedded into a pseudo-random pixel sequence.  Each of these operations requires that (at least one) secret key be pre-shared between sender and recipient, and anything else required for message extraction is considered additional data.  LSB embedding requires nothing beyond the pre-shared secret key, and this is true as well of the modification described in this paper.  The need for additional data, particularly data specific to individual stego images, increases the difficulty of practical implementation because it requires the existence of a secure channel that can be accessed on a per-message basis.

The powerful adaptive techniques described above work by selecting pixels for embedding such that some optimization criterion is achieved.  In order for the recipient to extract the message, they must know which pixels were embedded.  Many of these methods \cite{Holub,Holub2,Fridrich3,MiPOD,HILL} use syndrome trellis codes (STC) to reduce the number of pixel modifications, and require the parity-check matrix of the code for message extraction.  This matrix encodes the embedding specific to a single stego image, and so must be shared along with the image for extraction.  In addition, it must be kept secret since an adversary that intercepts it can use it to extract the message.  As an example, for HUGO \cite{Pevny} this matrix has dimensions $\sim \alpha n^2$ for an embedding rate of $\alpha$ into an $n$-pixel image.  For certain embedding rates, the data structure used to represent this matrix could be comparable in size to the message, in which case one might as well use the secure channel to exchange the message itself and forego steganography entirely.  In any case, the requirement that a secure channel be available for the exchange of secret data ``on-demand'' is potentially prohibitive for all but the most well-resourced of actors. 

Machine learning-based methods are also impractical from this standpoint, since they generally require that the recipient has a specially-trained neural network for message extraction.  The extractor must typically be trained on the same data set as the generator, and so in the above scenarios it is developed by the sender and must be sent to the recipient. Short of providing a fully-executable neural network, the sender could opt to send the recipient only the parameters (weights, biases, activations) of the network which they would then use to develop their own network.  The extractor networks, however, can be rather large and the data structure representing these parameters can be sizeable, {\it e.g.} almost 70 Mb for the model of \cite{GSIVAT} according to \cite{3player}. Some deep learning-based models that implement adaptive strategies \cite{SSGAN} or matrix embedding \cite{ADV-EMB} must additionally provide parity-check matrices.  And so, like the adaptive methods, machine learning-based models require considerable additional data for message extraction, challenging their practicality. 
\subsection{Prevalence of State-of-the-Art Steganalysis}
Deep learning has also been applied to steganalysis \cite{Xu-Net,Ye-Net,ReST-Net,Yedroudj-Net,SRNet,Chaumont,TanS}, notably as the discriminator networks in GAN-based steganography.  The 20-layer convolutional model, called Xu-Net \cite{Xu-Net}, serves as the discriminator for several of the above methods \cite{TangW,ASDL-GAN,UT-GAN}.  Outside of this application, deep learning-based steganalyzers are some of the most powerful general purpose detectors ever developed, with the 11-layer SRNet \cite{SRNet} cited as one of the most powerful at the end of 2018 \cite{Chaumont}.  While several of these models have publicly-available code \cite{SRNet,Yedroudj-Net,Ye-Net,Xu-Net}, as deep neural networks like the above GANs, considerable resources and expertise are required to train and implement these algorithms.  As noted for example in \cite{Chaumont}, SRNet ``requires strong know-how for its initialization.''  Meanwhile, machine learning-based steganalyzers without deep architectures, like rich models \cite{Kodovsky2}, and ensemble and SVM-based classifiers \cite{Kodovsky}, are available \cite{FridrichCode2,FridrichCode3} and easier to train, but still require careful hyperparameter optimization and regularization for good generalizability \cite{rich2}.   

Deep learning-based steganalysis is an exciting but immature technology, and its complexity and training requirements prevent its wide-spread adoption in practical detectors.   In contrast, the family of structural attacks, like SPA, Triples, and Weighted Stego, are publicly available, require no training, and work ``out-of-the-box''.  A low-resourced or unsophisticated ``warden'' might therefore be expected to opt for this brand of steganalysis, and it is this ``lower-risk channel'' for which a more secure LSB embedding algorithm, like ours and those reviewed in the Related Work section, might find useful application given its relative ease of use (in comparison with machine learning-based models) and ease of message extraction (unlike adaptive and machine learning-based models).

\section{Structural Steganalysis}
Structural steganalysis refers to a family of techniques that seek to detect hidden messages in spatial domain images by analyzing the statistical properties of contiguous groups of pixels.  These methods have had good success detecting randomized LSB embedding at even low embedding rates.  
\subsection{First-order Attacks}
First-order statistics, like frequency counts of pixel values, were the basis of the early histogram-based attacks.  Often referred to in the literature as the {\it histogram attack}, the approach of \cite{Westfeld1} employs a $\chi^2$ test to determine whether the tendency of LSB embedding to even-out the counts of consecutive even-odd pixel values can be distinguished from the histograms of typical cover images.  The histogram attack is particularly useful against serially embedded messages, but is only effective against randomized embedding when the relative payload is high, around 1 bit/pixel.  The trouble is that first-order statistics vary considerably from image to image, and so it is difficult to ascertain whether an image with nearly equal numbers of even-odd pixel values is hiding data, or whether it just looks that way naturally.  

Also based on the image histogram, the work of \cite{Harmsen} modeled LSB steganography as additive noise and observed that the smoothing-out of neighboring histogram bins observed in \cite{Westfeld1} could be quantified in terms of the center of mass of the histogram characteristic function.  In \cite{Harmsen}, this attack was only tested on a few color images at full embedding capacity, and so its performance against lower rates has not been carefully studied.  Absent good models of first-order statistics for natural images, we expect this method to likewise struggle to detect lower embedding rates.  
\subsection{Second-order Attacks}
Sample pairs analysis (SPA) \cite{Fridrich1,Dumitrescu1,Dumitrescu2,Ker1} considers the second-order statistics of natural images, and is based on the premise that natural images of objects with continuous shading should exhibit fairly small differences between neighboring pixels, and, for a given pair of such neighboring pixels, $(u,v)$, we should just as readily expect $u < v$ as $v < u$. This assumption is based on the expectation that natural images have no preferred direction of gradient.  It is further supposed that this parity between pairs with $u<v$ and $v<u$ should hold regardless of whether $u$ or $v$ happens to even or odd.  It can be shown that LSB embedding spoils this parity in a distinctive way, and SPA was developed to translate these observed parity deviations into an estimate of the hidden message length. 

\begin{figure}
\centering
\includegraphics[width=2in]{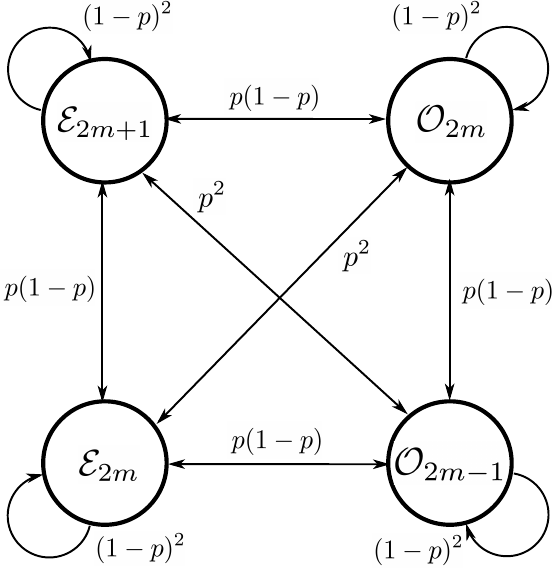}
\caption{\footnotesize{Transition probabilities of subsets of the trace set $\mathcal{C}_m$. }}
\label{SPA_trans}
\end{figure}

We define a {\it sample pair} as a doublet of neighboring pixels $(x_1,x_2)$, where each pixel (or channel for color images) takes on a $b$-bit value (typically $b=8$ bits).  All the pairs in an image form a multiset\footnote{A {\it multiset} is the generalization of a set to include non-unique elements.  Hereafter we will simply refer to them as sets.}, $\mathcal{P}$.  Interestingly, LSB embedding does not change the value $\floor{x_2/2} - \floor{x_1/2} = m$ of the pair $(x_1,x_2)$.  All such pairs form the {\it trace set}, 
\begin{equation}
\mathcal{C}_m = \{(x_1,x_2) \in \mathcal{P}\,|\,\floor{x_2/2} - \floor{x_1/2} = m\}.
\end{equation}
Further, a particular pair $(x_1,x_2)$ falls into one of two different subsets, where we use the concise notation of Ker \cite{Ker2}:
\begin{eqnarray}
\mathcal{E}_m &=& \{(x_1,x_2) \in \mathcal{P}\,|\,x_2 - x_1 = m, \,{\rm with}\, x_1 \,{\rm even}\},\\
\mathcal{O}_m &=& \{(x_1,x_2) \in \mathcal{P}\,|\,x_2 - x_1 = m, \,{\rm with}\, x_1 \,{\rm odd}\}.
\end{eqnarray}
The trace set $\mathcal{C}_m$ contains the four subsets: $\mathcal{E}_{2m}$, $\mathcal{O}_{2m-1}$, $\mathcal{E}_{2m+1}$, and $\mathcal{O}_{2m}$.  Now, the trace set $\mathcal{C}_m$ is {\it closed} under the action of LSB embedding, and so we expect the number of pairs in $\mathcal{C}_m$ of the cover to be the same as the number of pairs in the stego image,  $\mathcal{C}'_m$, that is,  $|\mathcal{C}_m|=| \mathcal{C}'_m|$, where vertical bars indicate the cardinality of the set.   However, the trace subsets are not closed under LSB embedding, with transitions occurring according to the diagram in Figure \ref{SPA_trans}.  The subsets transform under LSB embedding according to,
\begin{equation}
\left( \begin{array}{c}
E(|\mathcal{E}'_{2m}|)\\
E(|\mathcal{O}'_{2m-1}|)\\
E(|\mathcal{E}'_{2m+1}|)\\
E(|\mathcal{O}'_{2m}|) \end{array}\right) = \left(\begin{array}{cccc}
b^2 & ab & ab & a^2\\
ab & b^2 & a^2 & ab\\
ab & a^2 & b^2 & ab\\
a^2 & ab & ab & b^2\end{array}\right)\left(\begin{array}{c}
|\mathcal{E}_{2m}|\\
|\mathcal{O}_{2m-1}|\\
|\mathcal{E}_{2m+1}|\\
|\mathcal{O}_{2m}| \end{array} \right)
\label{LSB_trans}
\end{equation} 
where $a=p, b = 1 - p$, and $p$ is the probability that the LSB of a single pixel is changed.  The quantity $E(|\mathcal{E}'_{2m}|)$ is the expectation value of the cardinality of the set $\mathcal{E}'_{2m}$ after LSB embedding; it is a random variable because the embedding process is probabilistic.  In what follows, though, we will assume that the measured values of these sets are close to the expectations and simply write $|\cdot|$ in place of $E(|\cdot|)$. 

In terms of the above sets, the SPA\footnote{Often in the literature, the term ``SPA'' refers specifically to the technique used in \cite{Dumitrescu2} to compute the change rate, $p$, from Eq. (\ref{cr}); here, we use it more generally to refer to the $2^{\rm nd}$-order structural steganalysis and cover assumptions that yield Eq. (\ref{cr}.), irrespective of how it is solved.} cover image assumption can be written simply as $|\mathcal{E}_{2m+1}| = |\mathcal{O}_{2m+1}|$; it is clear from looking at Figure \ref{SPA_trans} that this condition will in general fail to hold under LSB embedding. 
For example, LSB embedding changes the cardinalities of the subsets $\mathcal{E}_{2m+1}$ and $\mathcal{O}_{2m+1}$ according to
\begin{eqnarray}
|\mathcal{E}'_{2m+1}| &=& p(1-p) |\mathcal{E}_{2m}| + p^2|\mathcal{O}_{2m-1}| \nonumber\\&+& (1-p)^2|\mathcal{E}_{2m+1}| + p(1-p)|\mathcal{O}_{2m}|,\\
|\mathcal{O}'_{2m+1}| &=& p^2 |\mathcal{E}_{2m+2}| + p(1-p)|\mathcal{O}_{2m+1}| \nonumber\\&+& p(1-p)|\mathcal{E}_{2m+3}| + (1-p)^2|\mathcal{O}_{2m+2}|.
\end{eqnarray}
In general, we therefore expect $|\mathcal{E}'_{2m+1}| \neq |\mathcal{O}'_{2m+1}|$.

To infer the embedding rate, $\alpha = 2p$, we must consider the inverse of Eq. (\ref{LSB_trans}) since the stego image only gives us access to the primed quantities, 
\begin{equation}
\left(\begin{array}{c}
|\mathcal{E}_{2m}|\\
|\mathcal{O}_{2m-1}|\\
|\mathcal{E}_{2m+1}|\\
|\mathcal{O}_{2m}| \end{array} \right) = \gamma\left( \begin{array}{cccc}
b^2 & -ab & -ab & a^2\\
-ab & b^2 & a^2 & -ab\\
-ab & a^2 & b^2 & -ab\\
a^2 & -ab & -ab & b^2\end{array} \right)\left( \begin{array}{c}
|\mathcal{E}'_{2m}|\\
|\mathcal{O}'_{2m-1}|\\
|\mathcal{E}'_{2m+1}|\\
|\mathcal{O}'_{2m}| \end{array} \right)
\label{LSB_Inv}
\end{equation} 
where $\gamma = (b-a)^{-2}$.  Under the assumption $|\mathcal{E}_{2m+1}| = |\mathcal{O}_{2m+1}|$, we obtain the quadratic expression,
\begin{eqnarray}
&&\alpha^2(|\mathcal{C}_m| - |\mathcal{C}_{m+1}|) +2\alpha(|\mathcal{E}'_{2m+2}|+|\mathcal{O}'_{2m+2}| - 2|\mathcal{E}'_{2m+1}| + 2|\mathcal{O}'_{2m+1}| \nonumber \\ &&-|\mathcal{E}'_{2m}|-|\mathcal{O}'_{2m}|)+4(|\mathcal{E}'_{2m+1}|-|\mathcal{O}'_{2m+1}|) = 0.
\label{cr}
\end{eqnarray}
There is an equation like this for each $m$, and to obtain $\alpha$ one option is to sum them all together and solve the resultant single quadratic equation.  Alternatively, the least squares method of \cite{Lu} can be used to estimate the value of $\alpha$ that minimizes the sum of the squared errors,
\begin{equation}
\hat{\alpha} = \underset{\alpha}{\arg \min} \sum_m (|\mathcal{E}_{2m+1}| - |\mathcal{O}_{2m+1}|)^2.
\end{equation}
Because this second approach generalizes well to alternative cover assumptions that we will be making, we adopt the least squares approach in this study.  

Sample pairs analysis using least squares optimization has proven quite successful at detecting embedding rates as low as 5\% \cite{Lu}, and the additional optimizations of \cite{Ker1} have achieved rates as low as 3\% \cite{Ker2}.  But, it is possible to do better by considering the higher-order statistics of larger sets of pixels.  
\subsection{Higher-order Attacks}
Ker \cite{Ker2} has developed a generalized approach for analyzing $n$-tuples of pixels; specifically, he explored whether the cardinalities of sets of triplets of consecutive pixels, $(x_1,x_2,x_3)$, can reveal LSB embedding.  Trace sets are defined in this case as 
\begin{equation}
\mathcal{C}_{m,n} =\left\{(x_1,x_{2},x_{3})\in \mathcal{P}\,|\,\floor{x_{i+1}/2} - \floor{x_i/2} = m_i\right\},
\end{equation}
with subsets
\begin{eqnarray}
\mathcal{E}_{m,n} &=&\left\{(x_1,x_{2},x_{3}) \in \mathcal{P}\,|\,x_{i+1} - x_i = m_i, \,{\rm with}\, x_i \,{\rm even}\right\},\\
\mathcal{O}_{m,n} &=& \left\{(x_1,x_{2},x_{3}) \in \mathcal{P}\,|\,x_{i+1} - x_i = m_i, \,{\rm with}\, x_i \,{\rm odd}\right\},
\end{eqnarray}
with $1\leq i \leq 2$.  Each trace set has eight subsets: $\mathcal{E}_{2m,2n}$, $\mathcal{O}_{2m-1,2n}$, $\mathcal{E}_{2m+1,2n-1}$, $\mathcal{O}_{2m,2n-1}$, $\mathcal{E}_{2m,2n+1}$, $\mathcal{O}_{2m-1,2n+1}$, $\mathcal{E}_{2m+1,2n}$, and $\mathcal{O}_{2m,2n}$. At higher-order, there are more symmetries to exploit for detection: there is the analog of the SPA parity symmetry, $|\mathcal{E}_{2m+1,2n+1}| = |\mathcal{O}_{2m+1,2n+1}|$, the order symmetry, $|\mathcal{E}_{m,n}| = |\mathcal{E}_{n,m}|$ for each $m,n$ (and likewise $|\mathcal{O}_{m,n}| = |\mathcal{O}_{n,m}|$),  and also a reflectional symmetry that relates sets under the transformation $(m,n) \rightarrow (-n,-m)$.  In \cite{Ker2}, only the parity symmetry is considered which leads to a cubic analog of Eq. (\ref{cr}) in terms of the variable $q = 1/(1-2p)$. This method, called {\it Triples analysis}, has some interesting properties: it performs rather unpredictably against stego images with high embedding rates ($\gtrsim 50\%$), but does well for {\it lower} rates.  It showed to be slightly more sensitive than RS and SPA, detecting embedding rates as low as 4\% and with a lower false alarm rate for uncompressed images; it did remarkably better than these lower-order techniques against JPEG-compressed images.  

Finally, an analysis of quadruples was also studied \cite{Ker3}.  The cover image symmetries considered were the analog parity symmetry, the inversion symmetry $|\mathcal{E}_{m,n,o}| = |\mathcal{O}_{-m,-n,-o}|$, and the permutative symmetry $|\mathcal{E}_{m,n,o}| = |\mathcal{E}_{\pi(m,n,o)}|$ for any permutation, $\pi$.  Each of these symmetries provides a separate estimate of the change rate, $p$, via a quartic polynomial in $q = 1/(1-2p)$.  These equations have multiple roots and it is not clear which one to choose as the best estimate for $p$: in \cite{Ker3}, Ker suggests selecting the root closest to a prior estimate of $p$ from SPA or Triples analysis.  When this can be done, the quadruples detector appears to be mostly consistent with lower-order tests.  
\section{Cover Modifications to Defeat SPA}
Sample pairs analysis is premised on the key assumption that natural images should satisfy the constraint $|\mathcal{E}_{2m+1}| \approx |\mathcal{O}_{2m+1}|$.  Certainly there are exceptions, but a decade's worth of analysis on a variety different image data sets confirms this hypothesis as generally true.  But, what if one could deliberately modify the statistics of the cover image to violate it prior to LSB embedding?  Would it be possible to modify the image in such a way that, after embedding a secret message, the statistics of the stego image are returned to those of the original cover?  This technique was recently demonstrated successfully in \cite{Shreelekshmi}, and we review it here in our own notation.  Hereafter, we refer to the practice of altering cover image statistics prior to applying steganography as {\it cover modification}.

\begin{figure}
\centering
\includegraphics[width=3.5in]{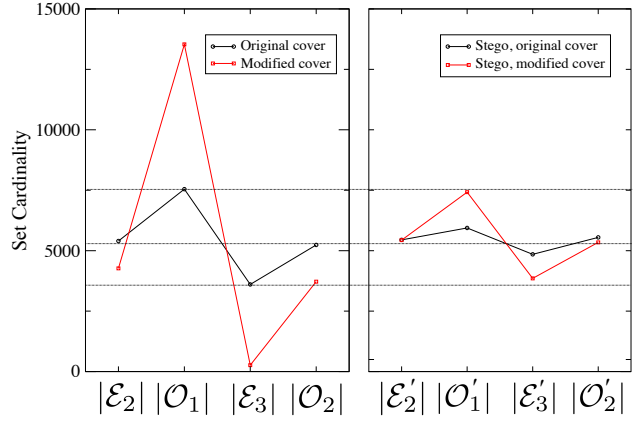}
\caption{\footnotesize{The effect of LSB embedding on trace subsets of an original and modified cover image.  On the left are subset cardinalities of the original and modified cover, and on the right is how these cardinalities change after some amount of LSB embedding.  The dashed horizontal lines are a guide to assess how well the ``stego modified cover'' (red) in the right plot resembles the ``modified cover'' (black) in the left plot.}}
\label{bars_demo}
\end{figure}

Schematically, LSB embedding transforms a cover image, $I$, into a stego image, $I'$, as $I' = L\cdot I$.  The basic idea is to come up with a transformation, $T$, such that $I' = LT\cdot I = I$.  While the pixel-wise act of embedding into LSB's cannot be inverted, we can invert the effects of LSB embedding on trace subset cardinality.  The desired transformation has already been written down: it is the matrix in Eq. (\ref{LSB_Inv}).  Here, though, we wish to apply this transformation not to the stego image trace sets, but the cover image sets,
\begin{equation}
\left(\begin{array}{c}
|\mathcal{E}^M_{2m}|\\
|\mathcal{O}^M_{2m-1}|\\
|\mathcal{E}^M_{2m+1}|\\
|\mathcal{O}^M_{2m}| \end{array} \right) = \gamma\left( \begin{array}{cccc}
b^2 & -ab & -ab & a^2\\
-ab & b^2 & a^2 & -ab\\
-ab & a^2 & b^2 & -ab\\
a^2 & -ab & -ab & b^2\end{array} \right)\left( \begin{array}{c}
|\mathcal{E}_{2m}|\\
|\mathcal{O}_{2m-1}|\\
|\mathcal{E}_{2m+1}|\\
|\mathcal{O}_{2m}| \end{array} \right)
\label{SPA_CM}
\end{equation} 
giving us {\it modified} trace sets, $\mathcal{C}^M_m$.  The effect on the $m=1$ trace subsets of a sample cover image is shown in Figure \ref{bars_demo}.  
Because $\mathcal{O}_1$ is larger than $\mathcal{E}_3$, Figure \ref{bars_demo} (left), there is a net transfer of pairs from $\mathcal{O}_1$ to $\mathcal{E}_3$ after LSB embedding (right).  But, if we anticipate these transitions by preemtively moving pairs from $\mathcal{E}_3$ to $\mathcal{O}_1$ prior to embedding, the stego image will exhibit the same statistics as the original cover, that is, the set cardinalities of trace subsets should be approximately equal (compare the black ``original cover'' in Figure \ref{bars_demo} (left) with the red ``stego modified cover'' on the right).  Sets like $\mathcal{E}_3$ from which pairs must be moved to other sets during cover modification are called {\it donor subsets}. 

The more data we wish to embed into the cover image, the more pairs need to be moved out of donor subsets.  Since the donor subsets are of finite cardinality, there is a limit to the embedding capacity that depends on the particular cover image.  Each trace set, $\mathcal{C}_m$, will have at least one donor subset, and the amount of data that can be embedded into pairs belonging to that trace set is constrained by the subset that empties at the smallest $\alpha$. In practice, to find this $\alpha$ we solve each equation of Eq. (\ref{SPA_CM}) separately with the left-hand-side set to zero (corresponding to an empty subset in the modified cover) and pick the smallest $\alpha$.  Call this $\alpha_m$, the smallest embedding capacity allowed by trace set $\mathcal{C}_m$.  Then, the maximum embedding capacity allowed for the image is the minimum capacity of all trace sets, $\alpha = \min \{\alpha_m\}$. 

The number of trace sets to modify is arbitrary, though good results are obtained for $-5\leq m \leq 5$.  This is the same range of sets found in \cite{Ker2} and \cite{Lu} to provide reliable detections, as higher-order trace sets tend to become sparsely populated and do not reliably satisfy the condition $|\mathcal{E}_{2m+1}| = |\mathcal{O}_{2m+1}|$.  Even within the lower-order sets, it might occur that one or a few trace sets severely constrain $\alpha$ such that their omission from embedding results in a higher embedding capacity.  As an extreme example, if a single donor set is already empty in any of the trace sets, the image cannot be embedded at all unless the pixels in this trace set are excluded from the embedding.  This is very uncommon for pairs analysis, though becomes more of a problem with higher-order cover modifications as we will see.  We therefore propose the following rule for identifying the maximum embedding capacity: set $\alpha = \alpha_{\tilde{m}}$, where, 
\begin{eqnarray}
\tilde{m} &=& \underset{m}{\arg \max}\, \alpha_m \left(1 - \frac{S_m}{N}\right) \\
S_m &=& \sum_{i|\alpha_i < \alpha_m} |\mathcal{C}_i|,
\label{max_rate}
\end{eqnarray}
where $N$ is the total number of pixels in the image.  We exclude all trace sets $\mathcal{C}_i$ with $\alpha_i < \alpha_{\tilde{m}}$ from the LSB embedding process.  The quantity $\alpha = \alpha_{\tilde{m}}$ is the effective maximum embedding rate that results after these trace sets have been excluded. 

In any case, once the value of $\alpha$ has been obtained, we are ready to perform the cover modification.  This is just a redistribution of pairs among the trace subsets according to Eq. (\ref{SPA_CM}) with the chosen embedding rate, $\alpha$.  The appropriate number of pairs are moved out of each donor subset into non-donor subsets according to their deficits.  For color images, trace sets are adjusted separately for each color channel.

\begin{figure}[ht]
\center
\includegraphics[width = 3.5in]{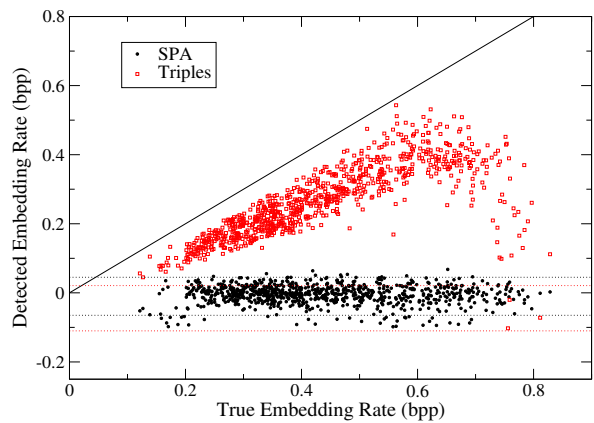}
\caption{\footnotesize{The results of SPA and Triples detections on 1000 uncompressed grayscale raster images with cover modifications made to defeat SPA.  Dashed lines indicate 95\% confidence limits for detection by the detector with the corresponding color.}}
\label{boss_couples}
\end{figure}

This kind of cover modification has been shown to be quite effective at evading SPA \cite{Shreelekshmi}, but what about higher-order attacks?  How does redistributing pixel pairs in this way affect the distribution of triplets?  We perform a test on 1000 512x512 images from the BOSS database\footnote{\texttt{http://agents.fel.cvut.cz/boss/}} of uncompressed, grayscale raster images.  LSB steganography was performed by embedding a pseudo-random bit stream, simulating an encrypted message, into pseudo-randomly selected LSBs at the embedding capacity of each image.  We present results in Figure \ref{boss_couples}: black points are the detected embedding rates using SPA, and the red squares are those using Triples.  The horizontal lines mark the 95\% confidence bounds for a detection with SPA (black) or Triples (red)\footnote{Confidence limits were established by running SPA and Triples detections on the raw un-embedded images.}.  Negative embedding rates are of course not possible, and are simply how these algorithms interpret certain set imbalances.  But, since negative $\alpha$ below the lower confidence bound might suggest that the image has been tampered with, such predictions can be considered detections.  Points that fall on the diagonal are perfect predictions of the true rate. Only cover images that are below the SPA detection threshold were selected for cover modification. 

The Triples analysis of \cite{Ker2} is able to detect the presence of a hidden message in almost every image, and estimate its length to within 50\% accuracy for most.  The noisiness observed in the Triples detections at high embedding rate possibly arises from the same instability observed by Ker in \cite{Ker2}.  And so, perhaps unsurprisingly, a second-order cover modification is insufficient for securing LSB embedding against higher-order structural attacks.   

\section{Cover Modifications to Defeat Both SPA and Triples}
To understand why the second-order cover modification did not also provide third-order protections, consider two consecutive triplets,
\begin{equation}
\overunderbraces{&\br{2}{m_1}& &\br{3}{m_3}& &\br{2}{m_5}}%
{(&x_1 , &x_2& , &x_3& | &x_4& , &x_5& , x_6&)}%
{& &\br{3}{m_2}& &\br{3}{m_4}},
\label{sextuplet}
\end{equation}
where the $m_i = x_{i+1}-x_i$ denote the differences of the indicated pixel values.  The pairs in this sextuplet belong to the trace sets $\mathcal{C}_{m_1}$, $\mathcal{C}_{m_3}$, and $\mathcal{C}_{m_5}$.   Embedding into this sextuplet will in general transform all the $m_i \rightarrow m_i'$; however, cover modifications based on pairs will only adjust pixels according to the transitions with $i = 1,3,5$.   But pixel $x_3$ also belongs to the first triplet, and so helps determine $m_2$, while pixel $x_4$ belongs to the second triplet and helps determine $m_4$.  If pixel $x_3$ is adjusted during cover modification according only to $m_3$, as would happen under a second-order cover modification, the effect on $m_2$ will be essentially random, and likewise for the effect on $m_4$ of adjusting $x_4$.  It would appear that in order to preserve both second- and third-order statistics after LSB embedding, we must make cover modifications at sixth-order, in terms of sextuplets\footnote{In what follows, we refer to an $n^{\rm th}$-order cover modification as one that adjusts the cardinalities of sets of $n$-tuples.}. 

\subsection{Attempt at a full sixth-order solution}
In \cite{Ker2}, Ker developed an approach to structural steganalysis to arbitrary order, which we apply here.  Trace sets carry five indices denoting the differences between consecutive pixels in the sextuplet,
\begin{eqnarray}
\mathcal{C}_{m_1,...,m_5} =\left\{(x_1,...,x_6)\in \mathcal{P}\,|\,\floor{x_{i+1}/2} - \floor{x_i/2} = m_i\right\}
\end{eqnarray}
and the subsets are defined analogously to the triplets case,
\begin{eqnarray}
\mathcal{E}_{m_1,...,m_5} &=&\left\{(x_1,...,x_6)) \in \mathcal{P}\,|\,x_{i+1} - x_i = m_i, \,{\rm with}\, x_i \,{\rm even}\right\},\\
\mathcal{O}_{m_1,...,m_5} &=& \left\{(x_1,...,x_6)) \in \mathcal{P}\,|\,x_{i+1} - x_i = m_i, \,{\rm with}\, x_i \,{\rm odd}\right\}.
\end{eqnarray}
There are 64 subsets in each $\mathcal{C}_{m_1,...,m_5}$ that can enumerated as follows \cite{Ker2}:  first, write $\mathcal{A}_{0,m_1,...,m_5}$ for $\mathcal{E}_{m_1,...,m_5}$ and $\mathcal{A}_{1,m_1,...,m_5}$ for $\mathcal{O}_{m_1,...,m_5}$.  Writing the concatenation of two sequences ${\bm s}$ and ${\bm t}$ as ${\bm s}.{\bm t}$, the trace subsets of $\mathcal{C}_{{\bm t}.k}$ can be obtained recursively from the subsets $\mathcal{A}_{\bm s_1}...\mathcal{A}_{\bm s_n}$ of $\mathcal{C}_{{\bm t}}$ as
\begin{eqnarray}
\mathcal{A}_{{\bm s_1}.(2k+\beta_1)},...,\mathcal{A}_{{\bm s_n}.(2k+\beta_n)},\nonumber \\
\mathcal{A}_{{\bm s_1}.(2k+\beta_1+1)},...,\mathcal{A}_{{\bm s_n}.(2k+\beta_n+1)}
\end{eqnarray}
where $\beta_i = 0$ if $\sum_i {\bm s_i}$ is even and $\beta_i = -1$ if the sum is odd.  Writing $P(\mathcal{A}_{{\bm s_i}},\mathcal{A}_{{\bm s_j}})$ for the transition probability between the two subsets $\mathcal{A}_{{\bm s_i}}$ and $\mathcal{A}_{{\bm s_j}}$ we have
\begin{eqnarray}
P(\mathcal{A}_{{\bm s_i}.(2k+\beta_i)},\mathcal{A}_{{\bm s_j}.(2k+\beta_j)}) &=& (1-p)P(\mathcal{A}_{{\bm s_i}},\mathcal{A}_{{\bm s_j}})\nonumber \\
P(\mathcal{A}_{{\bm s_i}.(2k+\beta_i+1)},\mathcal{A}_{{\bm s_j}.(2k+\beta_j)}) &=& pP(\mathcal{A}_{{\bm s_i}},\mathcal{A}_{{\bm s_j}})\nonumber \\
P(\mathcal{A}_{{\bm s_i}.(2k+\beta_i)},\mathcal{A}_{{\bm s_j}.(2k+\beta_j+1)}) &=& pP(\mathcal{A}_{{\bm s_i}},\mathcal{A}_{{\bm s_j}})\nonumber \\
P(\mathcal{A}_{{\bm s_i}.(2k+\beta_i+1)},\mathcal{A}_{{\bm s_j}.(2k+\beta_j+1)}) &=& (1-p)P(\mathcal{A}_{{\bm s_i}},\mathcal{A}_{{\bm s_j}})\nonumber\\.
\end{eqnarray}
Finally, the transition matrices can be obtained recursively from lower-order matrices via the $g$-fold Kronecker products,
\begin{eqnarray}
T_1 &=& \left(\begin{array}{cc}
1-p & p\\
p & 1-p\end{array} \right)
\\
T_{g+1} &=& \left( \begin{array}{c|c}
(1-p)T_g & pT_g\\
\hline
pT_g & (1-p)T_g\end{array}\right),
\end{eqnarray}
and similarly for the inverses,
\begin{eqnarray}
T^{-1}_1 &=& \frac{1}{1-2p}\left(\begin{array}{cc}
1-p & -p\\
-p & 1-p\end{array} \right)
\\
T^{-1}_{g+1} &=& \frac{1}{1-2p}\left( \begin{array}{c|c}
(1-p)T^{-1}_g & -pT^{-1}_g\\
\hline
-pT^{-1}_g & (1-p)T^{-1}_g\end{array}\right).
\end{eqnarray}

With so many more subsets per trace set at sixth-order, there is a real danger that we will encounter trace sets with at least one empty subset, preventing us from embedding into that trace set.  To find out, we test this cover modification on 1000 uncompressed grayscale raster images (512x512 PGM format from the BOSS database cited earlier) and 1000 uncompressed color raster images (high-resolution TIF format from the USDA NRCS archive\footnote{\texttt{http://photogallery.nrcs.usda.gov/}} resized to 640x450). 
Indeed, {\it all} of the trace sets in 20\% of the grayscale and 50\% of the color images had at least one empty subset, with the result that these images could not be modified and so could not serve as covers.  The remaining images in each data set overwhelmingly contained only a single trace set with no empty subsets (typically $\mathcal{C}_{0,0,0,0,0}$), with a resulting very low embedding capacity: 0.01\% and 0.02\% for grayscale and color, respectively.  Evidently, a full sixth-order cover modification can only be done at the expense of a virtually empty stego image.

\begin{figure*}[ht]
{\includegraphics[width = 5in]{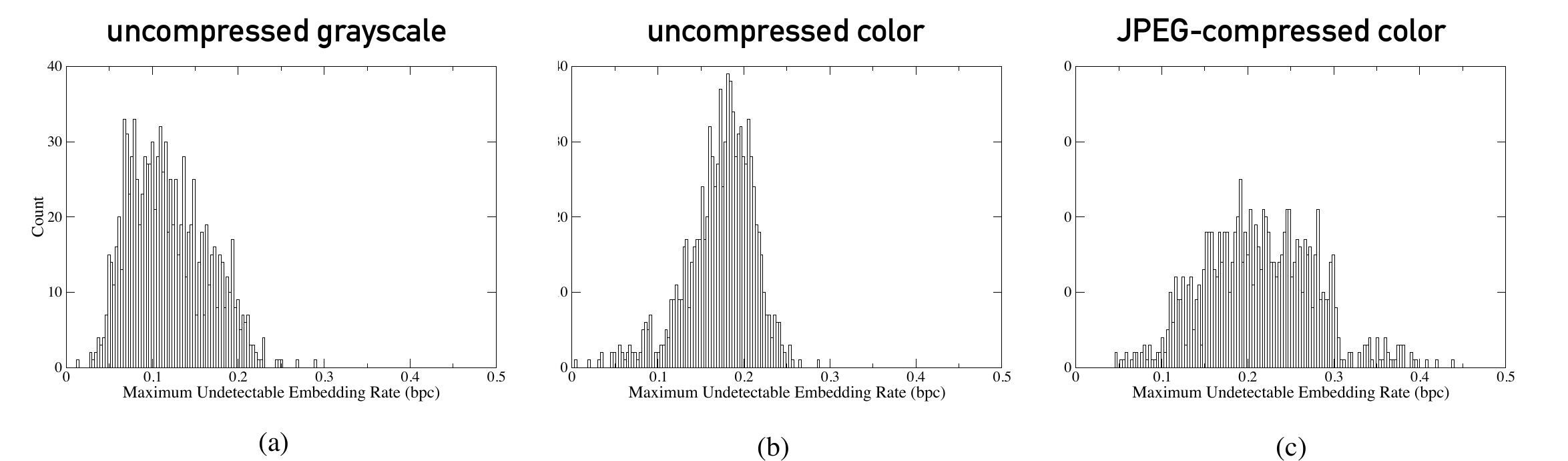}} 
\caption{\footnotesize{Distribution of maximum embedding rates $\alpha$ for images cover-modified to resist both SPA and Triples steganalysis. Results of 1000 (a) uncompressed grayscale, (b) uncompressed color, (c) JPEG-compressed color raster images.}}
\label{histos}
\end{figure*}
\subsection{Third-order solutions with partial embedding strategies}
The problem with the sixth-order approach is that a single empty subset excludes all the pixels in its trace set.  We cannot get rid of empty subsets because they are a property of the cover image which we intend to preserve; but, we can mitigate the collateral damage their exclusion has on other subsets.  One idea is to reduce the sizes of the trace sets so that the number of pixels that must be omitted from embedding is smaller in the event that the trace set contains an empty subset.  
One way to reduce the number of subsets per trace set is to reduce the dimensionality of the transformation:  while we are stuck preserving sixth-order statistics, we are not actually stuck with the sixth-order transformation, $T_6$, and the corresponding large 64-dimensional trace sets.  At the start of this section, we argued that the trouble with modifying second- or third-order statistics such that the other is also preserved lies in the middle pair, $(x_4,x_5)$, of the sextuplet Eq. (\ref{sextuplet}).  To accommodate these pixels, we were forced to consider the sixth-order statistics of the cover, which we've just seen is not possible.  Alternatively, we can simply {\it omit} these pixels from the embedding,
\begin{equation}
(x_1,x_2,\overbrace{\highlight{x_3|x_4}}^{{\rm omit}},x_5,x_6).
\label{middle}
\end{equation}
Then, $m_1$ and $m_2$ transform solely due to changes to pixels $x_1$ and $x_2$, and $m_4$ and $m_5$ solely to changes to pixels $x_5$ and $x_6$.  The middle index $m_3$ is left unchanged.  Cover modification can be done on triplets, and since only pairs that are fully inside triplets are embedded, these modifications will work to preserve second-order statistics as well.  In this way, we break the sixth-order problem up into two separate third-order problems.

Since only two pixels in each triplet are ever embedded, each triplet undergoes a second-order transformation governed by $T_2$ of Eq. (\ref{LSB_trans}).  The first triplet belongs to either the trace set $\{\mathcal{E}_{2m,2n}$, $\mathcal{O}_{2m-1,2n}$, $\mathcal{E}_{2m+1,2n-1}$, $\mathcal{O}_{2m,2n-1}\}$ or $\{\mathcal{E}_{2m,2n+1}$, $\mathcal{O}_{2m-1,2n+1}$, $\mathcal{E}_{2m+1,2n}$, $\mathcal{O}_{2m,2n}\}$, and the second triplet to either the trace set $\{\mathcal{E}_{2m,2n}$, $\mathcal{E}_{2m,2n-1}$, $\mathcal{E}_{2m+1,2n+1}$, $\mathcal{E}_{2m+1,2n}\}$ or $\{\mathcal{O}_{2m,2n+1}$, $\mathcal{O}_{2m-1,2n+1}$, $\mathcal{O}_{2m+1,2n}$, $\mathcal{O}_{2m,2n}\}$, and so the original 64-dimensional, sixth-order transition matrix decomposes as
\begin{equation}
\mathbf{64} = \mathbf{8}\otimes\mathbf{8} = (\mathbf{4}\oplus \mathbf{4})\otimes (\mathbf{4} \oplus \mathbf{4}).
\end{equation}
The four-dimensional trace sets are smaller and so the damage incurred from empty subsets is better contained. Subsets are also much less likely to be empty in the first place since they are larger and more inclusive. The cost is that we must exclude 1/3 of the image pixels from the embedding process\footnote{We also considered the strategy of omitting the four middle pixels from Eq. (\ref{sextuplet}), which, on its face would seem worse since twice as many pixels are excluded at the outset.  But, the trace sets are even smaller in this case and the subsets even more general, and so the maximum $\alpha$ could be large enough to compensate for the loss of pixels.  The LSB embedding transformation is composed of two separate families of single-pixel transformations on triplets: $\mathcal{E}_{2m,\Box} \leftrightarrow \mathcal{O}_{2m-1,\Box}$ with $\Box = 2n,2n+1$, and $\mathcal{O}_{2m,\Box} \leftrightarrow \mathcal{E}_{2m+1,\Box}$ with $\Box = 2n,2n-1$ for the first triple in each sextuplet (in which the first pixel is embedded); and $\mathcal{E}_{\Box,2n} \leftrightarrow \mathcal{E}_{2m,2n+1}$ with $\Box = 2m,2m+1$, and $\mathcal{O}_{2m,2n-1} \leftrightarrow \mathcal{O}_{\Box,2n}$ with $\Box = 2m,2m-1$ for the second triple in each sextuplet (in which the last pixel is embedded).  Each of these transitions defines a separate trace set, of which there are eight for each $m$ each with only two subsets: $\{\mathcal{E}_{2m,\Box},\mathcal{O}_{2m-1,\Box}\}$ and $\{\mathcal{E}_{2m+1,\Box},\mathcal{O}_{2m,\Box}\}$ for triplets $(x_1,x_2,x_3)$;  and $\{\mathcal{E}_{\Box,2m},\mathcal{E}_{2m,2n+1}\}$ and $\{\mathcal{O}_{2m,2n-1},\mathcal{O}_{\Box,2n}\}$ for triplets $(x_4,x_5,x_6)$.  The transition matrix decomposes as
\begin{equation}
\mathbf{64} = \mathbf{8}\otimes\mathbf{8} = \bigoplus_{i=1}^{4} \mathbf{2}_i \otimes \bigoplus_{i=1}^{4} \mathbf{2}_i ,
\end{equation}
so that each of the eight trace sets are acted on by $T_1$.  In comparison with the strategy in the body, though $\alpha$ tends to be larger for the image types tested, it does not offset the reduction in capacity from discarding the additional 1/3 of pixels.}. 

To perform this cover modification, we consider sets with $-5 \leq m,n  \leq 5$.  The first triplet in each sextuplet is placed into the set called $\mathcal{P}_1$ and the second into the set called $\mathcal{P}_2$.  The maximum embedding capacity is determined by applying Eq. (\ref{max_rate}) to $\mathcal{P}_1$ and $\mathcal{P}_2$ separately, and the smaller of the two is selected.  For almost all images, there were several omitted trace sets; all included trace subsets were then adjusted according to the method outlined in Section IV.  
We test the effectiveness of this cover modification against SPA and Triples analysis for the three image data sets: uncompressed grayscale, uncompressed color, and JPEG-compressed color raster images (\texttt{jpg} format from the NRCS archive resized to 640x450 and converted to bitmaps).
Only cover images that were consistent with zero embedding at 95\% CL according to both detectors were selected for testing.  Some images nonetheless still lead to detections at this threshold, and so for these we tuned the embedding rate down until it was no longer detected by either attack. The distribution of maximum embedding rates that escape both SPA and Triples detections are shown in Figure \ref{histos} for uncompressed grayscale images (a) and uncompressed color images (b).  Both images formats support undetectable embedding rates between around 0.05-0.25 bits per channel (bpc), with an average of 0.12 bpc for grayscale and 0.17 bpc for color. 

For JPEG-compressed images, we find that larger embedding rates are possible, with a range of 0.05-0.40 bpc, and an average of 0.21 bpc, Figure \ref{histos} (c).  This result is especially of interest since Triples analysis has shown to be much more reliable than pairs analysis at both detecting messages and estimating their length in JPEG-compressed covers, making it the last line of defense against these image types.  Cover modifications that resist these attacks at moderate embedding capacities might therefore be of considerable value. 

Before closing this section, we note that since first-order statistics, namely the quantities characterizing the distribution of single pixel values, are not adjusted during cover modification, the pixel value histogram will reflect LSB embedding.  However, the $\chi^2$ inference used to detect LSB embedding is not discriminating for randomly embedded messages with the relatively low embedding rates possible with cover modification at this order.
\subsection{Message Embedding and Extraction}
Once the cover modification is complete, messages can be embedded in the standard way, typically along a pixel path selected pseudo-randomly from the image.  This pseudo-random sequence can be generated via a stream cipher with a secret key shared between the sender and receiver.  The difficulty here, though, is that some pixels along the pseudo-random path might not be suitable for embedding for one of two reasons: i) the pixel belongs to an omitted trace set, or ii) the pixel belongs to the middle pair of a sextuplet (position $x_3$ or $x_4$ in Eq. (\ref{middle})), which must be excluded according to our chosen embedding strategy.  Assuming the sender performed the cover modification and so knows the omitted trace sets and the embedding strategy, they can simply skip these pixels when encountered along the pseudo-random path during the embedding procedure; the recipient, however, also must know which pixels have been skipped so that they extract data only from embedded pixels. Otherwise, the extracted message will contain point errors corresponding to non-embeddable pixels whose LSB's do not carry message bits.  Since the embedding strategy is a fixed feature of this method, it can be reasonably assumed that the recipient knows this strategy and so skips middle-pair pixels during extraction.  The recipient, however, does not in general know which trace sets have been omitted, because these depend on the particular cover image chosen, and so will unwittingly extract LSBs from these non-embeddable pixels.  This is not a problem as long as these LSBs can later be identified and removed.  
One way to do this is to simply provide the list of omitted trace sets to the recipient. For $-5 \leq m,n \leq 5$, this is a $2\times 11^2 = 242$ bit data structure, where the factor of two arises from the two separate families of trace sets: those of triplet type $(x_1,x_2,x_3)$ and those of type $(x_4,x_5,x_6)$.  This structure can typically be compressed by a factor of 5 or so, down to around 50 bits.  If a covert channel exists between the sender and receiver, this data can be directly shared.  A more convenient and practical solution is to embed it along with the message into the stego image: all that is needed is a contiguous group of embeddable pixels along the pseudo-random path large enough to contain the data structure (about 50 pixels for a 50 bit structure)\footnote{Even for the small 512x512 grayscale images considered in this study, such regions are typically plentiful with room to spare.}. The recipient first extracts the LSBs from the pixels along the pseudo-random path, skipping middle-pair pixels (but including the LSBs of pixels from omitted trace sets because these are unknown to the recipient.)  They then scan the extracted data for this 50-bit data structure: since it is embedded in a contiguous group of embeddable pixels, it can be recovered without error and used immediately to identify the omitted trace sets.  With this information, the recipient can then remove the LSBs of omitted pixels from the extracted data to obtain the correct message.

It is also standard to encrypt the message prior to embedding, both for confidentiality and so that, as an effectively pseudo-random bit sequence, the message won't introduce statistical artifacts into the image.  Here, if the message is encrypted and then embedded, the recipient's decrypted message will contain errors because the extracted sequence will contain additional bits---those corresponding to the LSBs of omitted pixels---that were ignored during encryption.  For example, schematically, key bit $k_i$ is used to encrypt message bit $m_{i}$.  But, suppose that there is a pair of pixels from an omitted trace set between the first and the $n^{th}$ pixels along the pseudo-random path.  Then, key bit $k_n$ will be used to decrypt message bit $m_{n-2}$, causing errors throughout the remainder of the sequence.  So, instead of encrypting the message itself, the sender must take into account the extra bits that the recipient will unwittingly attempt to decrypt along with the message.  Here is how that is done.   

Rather then encrypt the message itself (hereafter ${\bm m}_K$ for a message of length $K$), the sender encrypts a certain master sequence ${\bm M}$ that is constructed as follows.  First, the sender computes the pseudo-random pixel path, ${\bm x}$, through the image, excluding middle-pair pixels, but including pixels from omitted sets.   If pixel $x_i$ is embeddable, then $M_i = m_i$; else, set $M_i$ to some arbitrary constant, say $M_i = 1$.   These values are arbitrary because they are merely serving as place holders to keep the encryption and decryption processes in synch; these values are never actually used in the embedding.  The master sequence, ${\bm M}$, therefore includes both message bits and place holder values (corresponding to omitted pixels) in the order determined by the pseudo-random pixel path. Then, ${\bm M}$ is encrypted bit-wise to form the sequence $\overline{{\bm M}}$, and embedded along the pseudo-random pixel path as follows: if $\overline{M}_i$ is a message bit, embed it into the LSB of $x_i$; otherwise, skip it and the pixel $x_i$.  

\begin{figure}
\center
{\includegraphics[width = 3.5in]{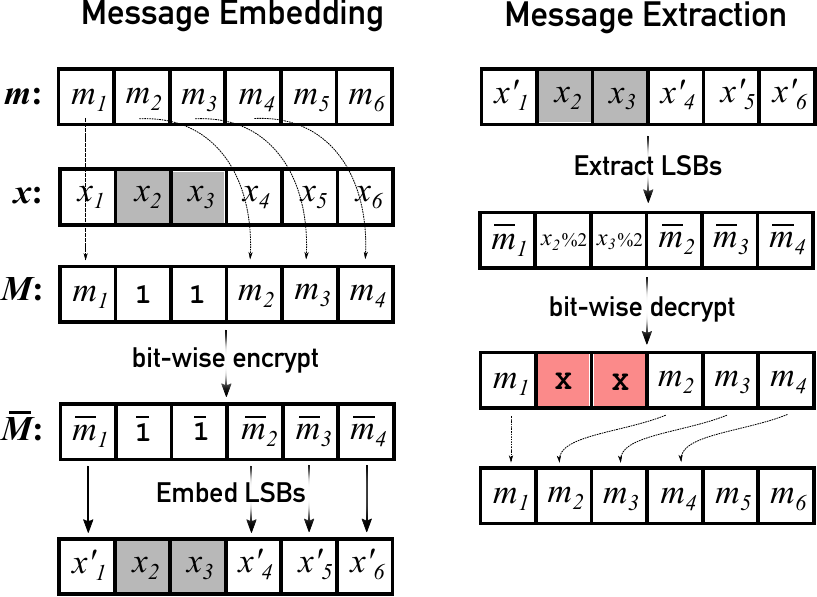}}
\caption{\footnotesize{Message encryption/embedding (left) and extraction/decryption (right) processes.  Gray squares indicate omitted pixels and red squares indicate decryption errors. See text for symbol definitions and discussion.}}
\label{enc_emb}
\end{figure}
Upon receipt, the recipient extracts all LSBs along the pseudo-random path (excluding middle-pair pixels, since the embedding strategy is assumed known to them), and decrypts the resulting bit sequence obtaining the sequence $\widetilde{{\bm M}}$. 
The LSBs of pixels from omitted trace sets that were not embedded will generally decrypt in error (won't decrypt to the place holder value), and so ${\bm M} \neq \widetilde{{\bm M}}$, but as long as the cipher is synchronous these will be point errors and not affect neighboring bits.  Therefore, message bits will decrypt correctly, with $\widetilde{m}_i = m_i$.  The recipient then scans the decrypted sequence for the 50-bit data structure indicating the omitted trace sets, and uses this information to drop the associated LSBs to recover the message, ${\bm m}$.  This process is illustrated schematically in Figure \ref{enc_emb}.   
\section{Security against higher-order structural attacks}
We have demonstrated that LSB embedding can be secured against second- and third-order structural attacks.  But this raises the obvious question: is it susceptible to fourth-order attacks?  In principle, yes.  The quadruples analysis of \cite{Ker3} was shown generally effective but difficult to apply, owing to the uncertainty over which root of the quartic polynomial for $q$ to select as the predicted change rate.  Ker suggests selecting the root closest to the estimate from a prior detection using SPA or Triples; however, these methods fail to detect any embedded message for covers modified according to Section V.  It is therefore unclear how quadruples analysis could be applied in practice against these kinds of stego images.  

An extension of this methodology to provide fourth-order protections is possible, but we find that embedding capacities are close to zero.  This is due to two factors: the loss of available pixels from the embedding strategy, and the limits imposed by donor set cardinality during cover modification.  The embedding strategy is necessary since to preserve all statistics up to fourth-order,  one needs to work with $4\times 3 = 12$-tuples of pixels and these large trace sets are almost guaranteed too all have at least one empty subset (easily as big a problem as with the sixth-order cover modification seen earlier).  The embedding strategy is similar to that considered in Section V: at $12^{\rm th}$-order, there are no pairs straddling quadruples as there were at third-order (the ``middle pair''), but there are triplets which would need to be omitted from the embedding for the same reasons,
\begin{equation}
(x_1,x_2,x_3,\highlight{x_4|x_5,x_6,x_7,x_8|x_9},x_{10},x_{11},x_{12}).
\end{equation}
This strategy reduces the number of available pixels to $6/12 = 1/2$ the total.  Average embedding capacities for a fourth-order cover modification are around 8\% of embeddable pixels, giving a total capacity for the image of around 4\%.  This is likely too low to be of any practical use. 

In general, the order of the cover modification is the least common multiple of all relevant orders whose statistics are to be preserved.  Let the highest-order preserved statistic be $k$, and let the least common multiple be $n$.  Then, only $\floor{k/2}+1$ pixels in the first and last $k$-tuples in each $n$-tuple are embeddable (all the others belonging to tuples that straddle the interior $k$-tuples.)  The result is that only the fraction $2\times(\floor{k/2} +1)/n$ are embeddable pixels for a general $n^{\rm th}$-order cover modification.  For $k=5$, this fraction is 1/10 and so on average overall embedding capacities are less than 1\%. We conclude from this that practical protection against structural steganalysis via cover modification does not extend beyond $k=3$.  
\section{Conclusions} 
This work has explored the extent to which LSB embedding can be made secure against structural steganalysis by modifying consecutive pixel count statistics of cover images prior to message embedding.  It is observed that modifications to protect against structural steganalysis at a particular order do not secure LSB embedding against higher-order attacks.   Given the effectiveness of the third-order Triples analysis of \cite{Ker2} at detecting moderate LSB embedding rates, particularly against JPEG-compressed images, we sought in this research to develop a cover modification that would be protective against both Sample Pairs and Triples analyses.  

We found that the sixth-order cover modification necessary to preserve both the second- and third-order cover statistics targeted by Sample Pairs and Triples analyses resulted in virtually zero embedding capacity.  This is because the large, 64-dimensional trace sets overwhelmingly tend to have at least one empty subset, preventing the redistribution of sextuplets within that trace set.  We therefore considered instead reverting to a third-order cover modification, but only embedding into certain pixels so that both second- and third-order statistics would be preserved.  Specifically, if all but the middle two pixels in each sextuplet are available for embedding, redistribution of pixel triplets also preserves second-order statistics and moderate embedding rates can be achieved.  We find that for uncompressed color and grayscale raster images,  undetectable embedding rates range from around 0.05-0.30 bpc, with an average of 0.12 bpc and 0.17 bpc, respectively.  For JPEG-compressed color images, we find generally higher undetectable payloads upwards to 0.40 bpc, with an average of 0.21 bpc.  Since Triples analysis has shown to be superior to SPA at detecting the presence of messages and estimating their length in JPEG-compressed images \cite{Ker2}, cover modifications that can defeat Triples are especially salient for this image type. 

We also conclude that cover modifications performed at higher than third order result in virtually zero embedding capacity, and so protections cannot be extended beyond Triples analysis.  This finding suggests that quadruples and even higher-order structural steganalysis should continue to be matured and developed in the face of these kinds of cover modifications.

Though accurate and powerful, structural steganalysis is not the only attack against LSB embedding. For example, the weighted stego-image \cite{Fridrich2,Ker5} and asymptotic uniform most powerful (AUMP) \cite{Fillatre} tests are robust detectors of LSB embedding that operate according to different principles, and so are not defeated with these kinds of cover modifications.  It is an open question whether the cover statistics targeted by structural steganalysis can be modified while also preserving the cover models exploited by weighted stego-image and AUMP steganalysis. Our approach might also be extended to secure against more general pixel grouping geometries like those explored in the Closure of Sets work of \cite{Khosravirad,Khosravirad2}.

Lastly, a nagging shortcoming of this methodology is the need to omit the pixels of entire trace sets in order to increase the maximum embedding rate.  This requires that the recipient perform the additional work of identifying and removing the LSBs of the omitted pixels from the extracted data before a meaningful message can be recovered.  Further, of course, having these pixels available for embedding in the first place would considerably increase the embedding capacity in many cases.  Future work could explore cover pre-processing (prior to the modifications studied here) that redistributes pixels in trace sets with small (and, hence, limiting) donor subsets; such transfers, however, would not be reversed in the course of LSB embedding and so would stand as permanent modifications to the cover image.  Such alterations would need to be performed carefully to avoid the introduction of statistical artifacts, and hence warrant further study.  

\section*{Acknowledgment}
The author thanks colleague Max Kresch for helpful discussions and for providing the NRCS image data set.  The author acknowledges use of the software available at \texttt{http://dde.binghamton.edu/}.

\bibliographystyle{ws-jcsc}
\bibliography{powell_rev2}
\end{document}